\documentclass[twocolumn,prb]{revtex4}
\usepackage{graphicx}    
\usepackage{dcolumn}     
\usepackage{slashbox}     
\usepackage{bm}          

\begin{document}
\title{Possibility of spin fluctuation mediated $d+id'$ pairing in 
a doped band insulator $\beta$-MNCl (M=Hf,Zr) superconductors}

\author{Kazuhiko Kuroki}

\affiliation{
Department of Applied Physics and Chemistry, 
The University of Electro -Communications, Chofu, Tokyo 182-8585, Japan
}

\begin{abstract}
We study two types of models for the superconducting 
layered nitride $\beta$-MNCl(M=Hf,Zr); 
a single band model on a triangular lattice, and a two band model 
on a honeycomb lattice. We find that the former model does not suffice as 
an effecitive model, while the latter one can be a good candidate.
We propose from the study on the two band model 
a possibility of spin fluctuation 
mediated $d+id'$-wave superconductivity in the doped $\beta$-MNCl.
We show that the relatively high $T_c$ obtained in the doped 
band insulator is a characteristic feature of the spin fluctuation 
mediated superconductivity on a honeycomb lattice. 
We also find that the gap anisotropy on the Fermi surface 
strongly increases upon increasing the doping concentration, and the 
inter layer hopping suppresses superconductivity. These results are 
in qualitative agreement with the experimental findings.
\end{abstract}
\pacs{PACS numbers: }
\maketitle

\section{Introduction}
Layered nitride $\beta$-$M$NCl\cite{Yamanaka} ($M=$Hf, Zr) doped with 
carriers is one of the most interesting group of superconductors.
The mother compound 
$\beta$-$M$NCl is composed of alternate stacking of honeycomb (HC) 
$M$N bilayer and Cl bilayer.\cite{Shamoto}
This is a band insulator and becomes a superconductor 
upon doping electrons by Na or Li intercalation.
They exhibit relatively high $T_c$ 
up to $\sim 25$K for $M=$Hf and $\sim 15$K for $M=$Zr.
The bilayer HC lattice structure consisting of $M$ and N 
is considered to be playing the 
main role in the occurrence of superconductivity, and the two 
dimensional nature of the superconductivity has been revealed by 
nuclear magnetic resonance (NMR)\cite{Tou3} 
and muon spin relaxation ($\mu$SR) studies.\cite{Uemura2,Uemura} 

Despite the relatively high $T_c$, 
experimental as well as theoretical studies indicate 
extremely low density of states (DOS)
at the Fermi level.\cite{Tou,Taguchi,Weht} In fact, 
they have the highest $T_c$ 
among materials with the specific heat coefficient $\gamma$ as small as 
$\sim$ 1mJ/molK$^2$.
The electron phonon coupling is also 
estimated to be weak,\cite{Tou,Weht,Taguchi2,Heid} and the isotope effect 
is found to be small.\cite{Tou2,Taguchi4} These experiments suggest that 
some kind of unconventional pairing mechanism may be at work, 
but on the other hand, 
NMR knight shift measurement suggests spin singlet pairing,\cite{Tou5}
and the tunneling spectroscopy\cite{Ekino}  and specific heat\cite{Taguchi} 
experiments find a  fully open, seemingly $s$-wave like gap.
However, regarding this fully open gap, recent experiments 
show that the anisotropy of the gap increases with doping. 
Namely, when the doping concentration is small, 
the specific heat coefficient increases linearly as 
a function of the magnetic field, suggesting an isotropic gap,\cite{Kasahara} 
while a steep increase of the coefficient at low magnetic field 
is observed for higher doping.\cite{Taguchi,Kasahara}
Also, $2\Delta/(k_BT_c)$ determined from the specific heat measurement 
ranges from $\sim 5$ in the lightly doped regime to less than $\sim 3$ in the 
heavily doped regime,\cite{Kasahara} which may be an indication that 
the maximum and minimum values of the gap have a  
substantial difference when heavily doped, 
and the specific heat is mainly governed by the 
minimum value.  Moreover, 
a recent superfluid density measurement by $\mu$SR 
shows that the gap is nearly 
isotropic when the doping concentration is small, while the 
anisotropy increases for large doping.\cite{Hiraishi}
Very recently, absence of coherence peak in the 
spin lattice relaxation rate has been found in an NMR experiment,\cite{Tou4}
again suggesting unconventional pairing, most probably with some kind of 
sign change in the superconducting gap.
Furthermore, for Li$_x$HfNCl, an intercalation of organic molecules 
tetrahydrofuran (THF) between the layers is found to enhance $T_c$.
\cite{Takano} 

Given these experimental circumstances,
here we study two types of models for the doped $\beta$-MNCl, 
a two band model on a HC lattice, and a single band model where the 
nitrogen site is effectively integrated out. While the latter 
model does not seem to suffice for explaining the 
relatively high $T_c$ in this series of material, 
the former one provides a possibility of spin fluctuation 
mediated $d$-wave superconductivity in the doped $\beta$-MNCl.\cite{Bill} 
Spin fluctuation mediated pairing in a doped {\it band insulator} 
may sound surprising and odd at first sight since one cannot expect 
strong spin fluctuations. 
Nevertheless, we show that the relatively high $T_c$ despite the low DOS 
and the temperature independent spin susceptibility 
is a characteristic feature of the spin fluctuation 
mediated superconductivity of a doped band insulator on the 
HC lattice. 
The most probable pairing state below $T_c$ is the $d+id'$ state, whose 
gap anisotropy on the Fermi surface is found to be 
strongly enhanced upon increasing the carrier concentration.
We also study the effect of the three dimensionality, where the 
inter layer hopping suppresses superconductivity. 

\section{Formulation}
\begin{figure}[h]
\begin{center}
\includegraphics[width=6cm]{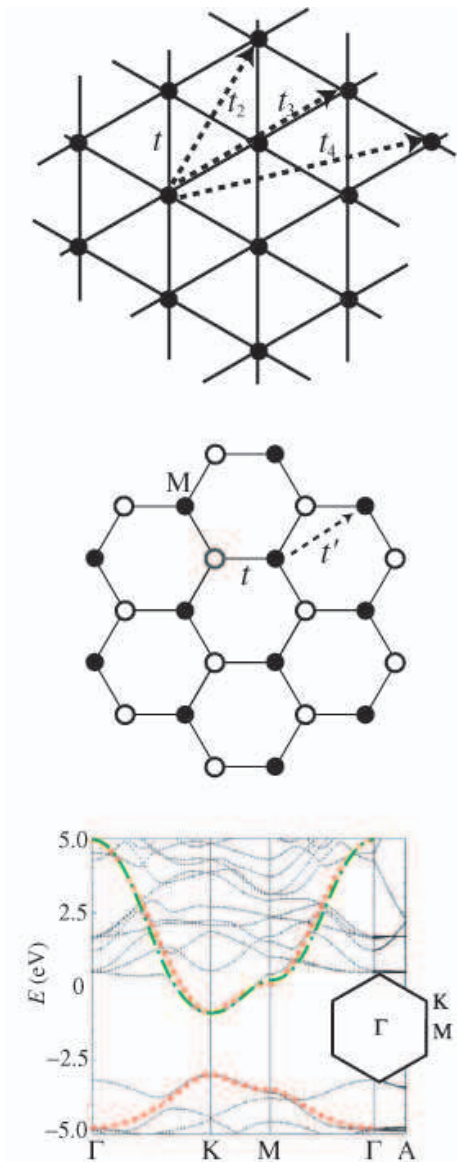}
\caption{\label{fig1} (color online) 
Upper :  the single band model on a triangular lattice.
Middle : the two band model on a honeycomb lattice.
Lower : the first principles band calculation taken 
from ref.\cite{Weht} and the band dispersion of the single band model 
(dash-dotted green) and the two band model(dashed red).}
\end{center}
\end{figure}

We first obtain the effective model for $\beta$-MNCl. 
Although the material consists of bilayer HC lattice, here we
consider models on a single layer lattice, 
and try to reproduce the bands that lie close to the 
Fermi level in the first principles band calculations.
The simplest one is a single band model on a triangular 
lattice, in which the 
nitrogen site degrees of freedom is effectively integrated out, 
so that there is only one site per unit cell.
This is a model that corresponds to the 
single band Hubbard model for the cuprates, where the oxygen $2p_x$, 
$2p_y$ degrees of freedom is integrated out from the original 
three band $d$-$p$ model.
By considering hopping integrals up to fourth nearest neighbors 
(Fig.\ref{fig1}, upper panel), we can reproduce the single band 
that crosses the Fermi level in the 
first principles band calculations.\cite{Weht,Heid,Hase,Oguchi}  
The band dispersion of the single band model with 
$t=0.65$ (eV), $t_2/t=-0.03$, $t_3/t=0.06$, $t_4/t=-0.025$ 
is shown by the green dash-dotted line in Fig.\ref{fig1}. 

To go a step further, we also consider a model 
on a single layer HC lattice 
consisting of alternating ''M'' and ''N'' 
orbitals with a level offset, which  roughly reproduces  
the conduction and valence bands closest to the Fermi level, 
as shown by the dashed lines in the middle panel of Fig.\ref{fig1}. 
Here we take the nearest neighbor (M-N) hopping 
$t=1.2$eV, the level offset $\Delta/t=2.7$, 
and the next nearest neighbor (M-M) hopping $t'/t=0.35$. 

The on-site interaction $U/t=6$ is considered in the single band model, 
and the on-site interaction $U/t=6$ is introduced   
on both M and N orbitals in the two orbital model\cite{comment}. 
The band filling $n$ is defined as the number of electrons/number of sites.
For the single band model, $n=2x$, where $x$ is the Li or Na content, since 
there are two M sites per unit cell, which results in 
bonding and antibonding bands, and the electrons are doped only 
in the bonding band. For the two band model, 
$n=1$ corresponds to the non-doped case, 
and $x$ and $n$ is related by 
$n=1+x$. We use fluctuation exchange (FLEX) method
\cite{Bickers,Dahm,Grabowski},
which is kind of a self-consistent random phase approximation,
to obtain the Green's function.
Then we solve the linearized Eliashberg equation,
\begin{eqnarray*}
\lambda\phi_{l m}(k)\nonumber &=& -\frac{T}{N}\sum_{k'}\sum_{l',m'}\\
&&\times V_{l m}(k-k')G_{ll'}(k')G_{mm'}(-k')\phi_{l'm'}(k').
\end{eqnarray*}
Here, $G$ is the Green's function matrix 
(with $l,m,$etc... labeling sites in a unit cell) obtained by FLEX.
$V$ is the spin singlet pairing interaction matrix 
given by $V=\frac{3}{2}U^2\chi_{\rm sp}-\frac{1}{2}U^2\chi_{\rm ch}$ 
with the spin and charge susceptibility matrices 
$\chi_{\rm sp(ch)}(q)=\chi_{\rm irr}(q)[1-(+)U\chi_{\rm irr}(q)]^{-1}$,
where $\chi_{\rm irr}(q)$ is the irreducible susceptibility matrix 
$\chi_{\rm irr}(q)= -\frac{1}{N}\sum_k G(k+q)G(k)$ 
($N$ is the number of $k$-point meshes).
In the following, the maximum eigenvalue of the spin susceptibility 
matrix will be referred to as the spin susceptibility.
We take up to $128\times 128$ $k$-point 
meshes and up to 16384 Matsubara frequencies. 
The eigenvalue of the Eliashberg equation $\lambda$ 
increases upon lowering the temperature and reaches unity at $T=T_c$.

\section{Single band model: Calculation results}
\begin{figure}[t]
\begin{center}
\includegraphics[width=8cm]{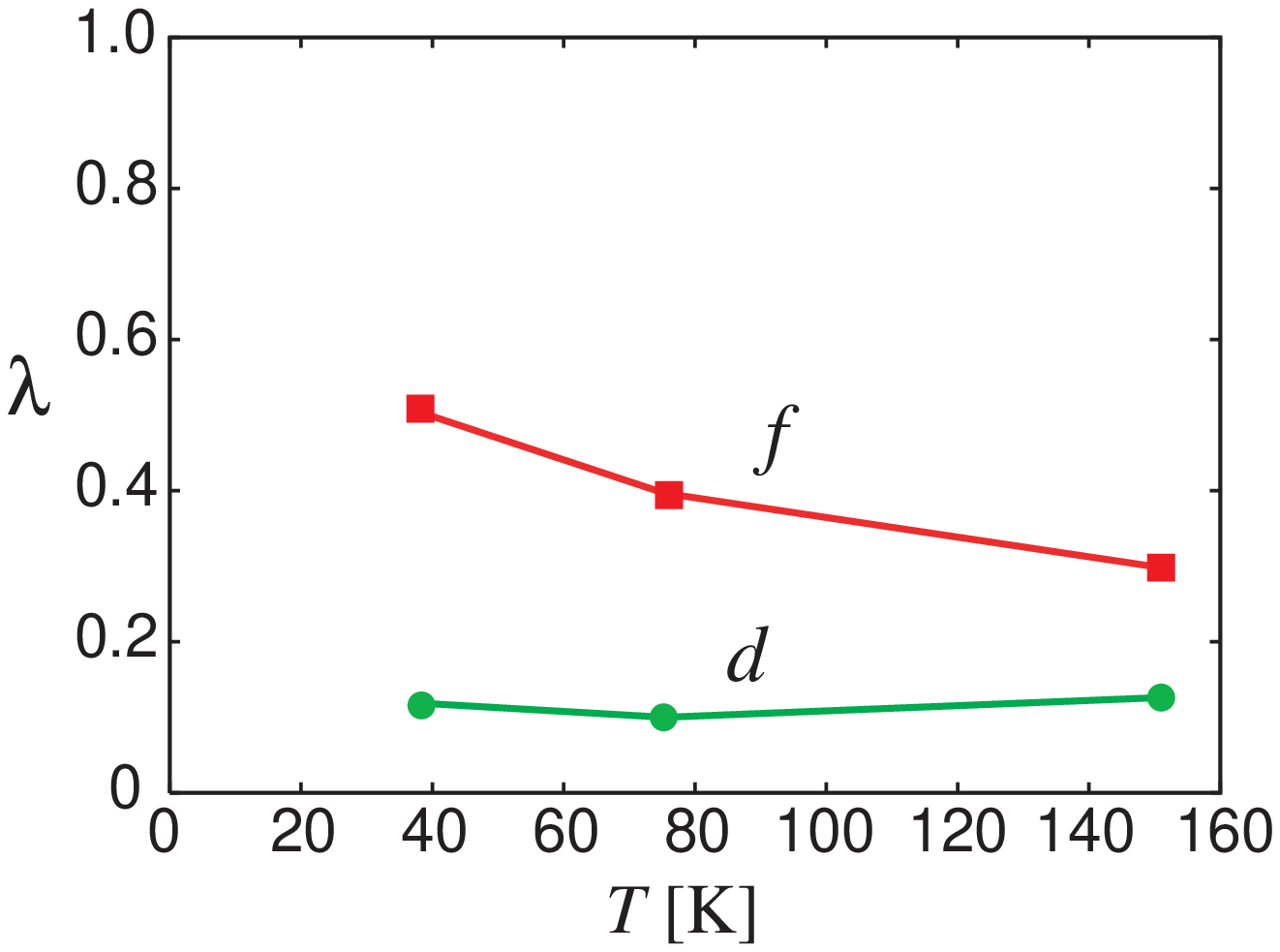}
\caption{\label{fig2} (color online) Eigenvalue of the Eliashberg 
equation as functions of temperature for $f$-wave and $d$-wave
pairings in the single band model.}
\end{center}
\begin{center}
\includegraphics[width=8cm]{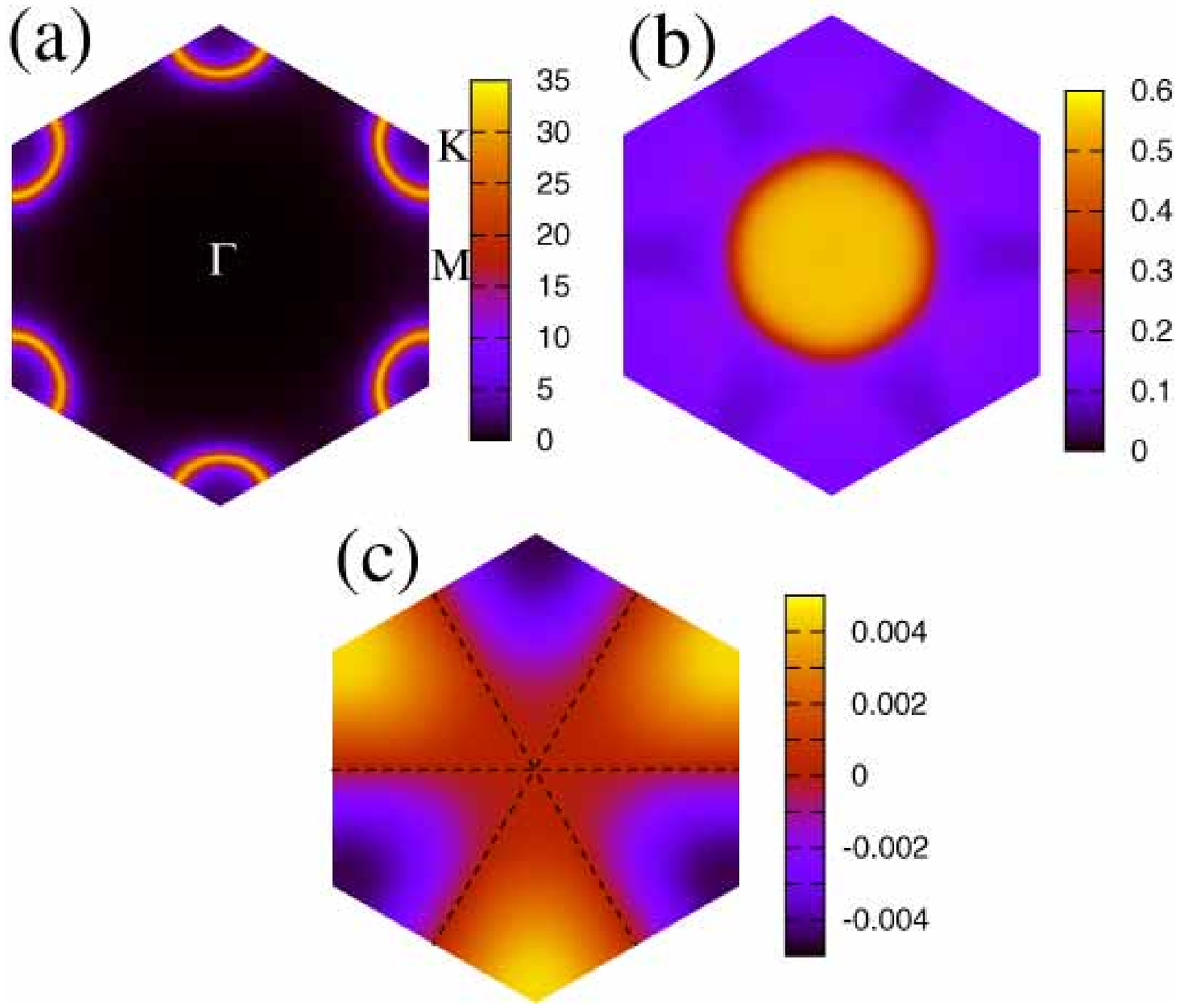}
\caption{\label{fig3} (color online) The contour plots of the FLEX result 
at the lowest Matsubara frequency for 
the single band model in the hexagonal Brillouin zone
with $n=0.2$ and $T=0.02t$ (a) The 
Green's function of the upper band squared,  
(b) the spin susceptibility, (c) the superconducting gap.}
\end{center}
\end{figure}

We first present calculation results for the single band model.
In Fig.\ref{fig2}, the eigenvalue of the Eliashberg equation is shown 
as functions of temperature for spin singlet and triplet pairings for 
$n=0.2$.
It can be seen that the triplet pairing strongly dominates over 
singlet pairing. The gap of the spin triplet pairing has 
an $f$-wave form, where the nodes of the gap 
do not intersect the Fermi surface (the ridge of the Green's function 
squared) as shown in Fig.\ref{fig3}. 
On the other hand, the singlet pairing has a $d$-wave gap (not shown).
The possibility of this type of spin triplet $f$-wave 
pairing was in fact proposed on a canonical triangular 
lattice in ref.\cite{KKA}. The triplet pairing dominates over singlet 
pairing because ferromagnetic spin fluctuations arise due to the dilute 
band filling as shown in Fig.\ref{fig3}(b), 
and also because the nodes of the $f$-wave gap do not 
intersect the Fermi surface.

Nonetheless, as can be seen from Fig.\ref{fig2}, 
the temperature at which the $f$-wave eigenvalue reaches 
unity, if any, seems to be very low. We have also tried a model 
that considers the nearest neighbor off-site interaction $V$ in addition to 
$U$. In this case, strong charge fluctuations arise, which makes the 
 triplet vs. singlet competition  more 
subtle, but in any case, the eigenvalue of the Eliashberg equation 
remained small even when the magnitude of $V$ is close to the 
point where the charge susceptibility diverges.
Considering the experimental fact that the $T_c$ is relatively 
high in the doped $\beta$-MNCl,  
and also that the pairing occurs in the spin singlet channel\cite{Tou5}, 
we believe that the single band model does not suffice 
as an effective model for $\beta$-MNCl.

\section{Two band model : Calculation results}
\subsection{Superconducting transition temperature}
\begin{figure}[h]
\begin{center}
\includegraphics[width=7cm]{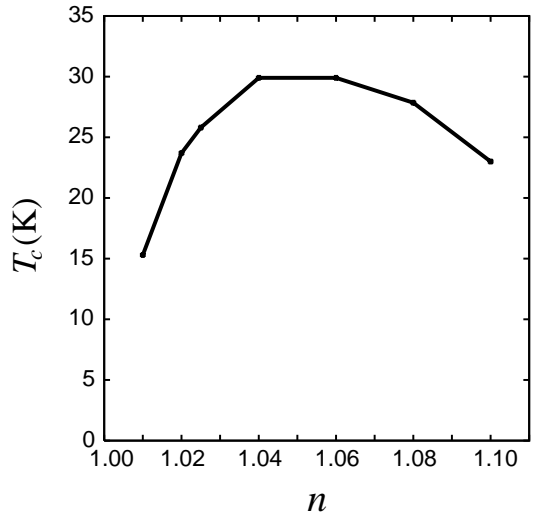}
\caption{\label{fig4} $T_c$ plotted as a function of the band filling for the 
two band model on the HC lattice.}
\end{center}
\end{figure}

We now move on to the two band model. 
In this model, we find that spin singlet pairing strongly 
dominates over triplet pairing.
In Fig.\ref{fig4}, we plot $T_c$ of the singlet pairing superconductivity 
as a function of the band filling. 
It can be seen that $T_c\sim 30$K is obtained, which can be considered as 
relatively high noting that $T_c\sim 100$K (assuming $t\sim 0.4$eV 
appropriate for the cuprates) is obtained by the same method 
for the Hubbard model on the square lattice.\cite{Bickers} 
The difference between the 
single and two band models is dramatic in that not only the 
$T_c$ but even the leading pairing symmetry is different.
This is in contrast to the case of the 
cuprates, where the single band Hubbard and three band $d$-$p$ models 
is expected to give roughly the same conclusions. 

\begin{figure}
\begin{center}
\includegraphics[width=8cm]{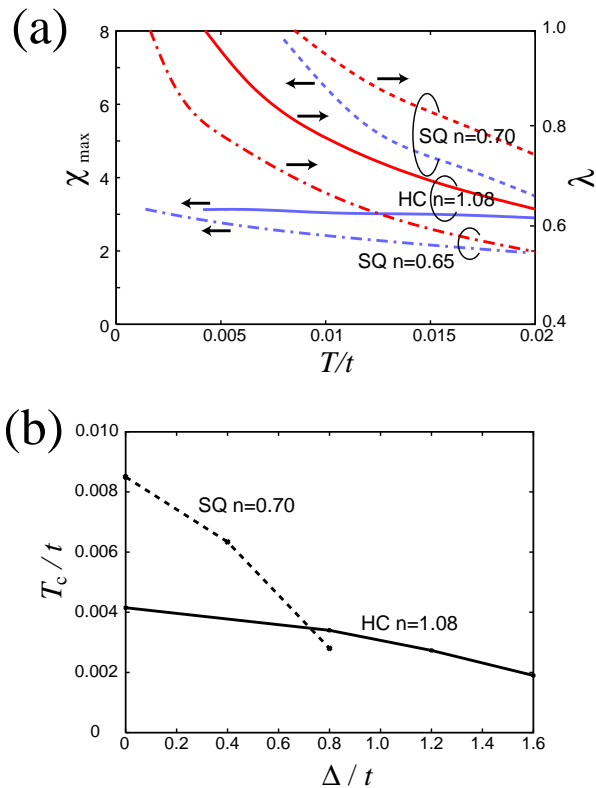}
\caption{\label{fig5} (color online) 
(a) The maximum value of the spin susceptibility 
(blue, left axis) and the eigenvalue of the 
Eliashberg equation (red, right)
as functions of temperature 
for the square lattice with $n=0.7$ (dashed) or $n=0.65$ (dash dotted) and 
for the honeycomb lattice (solid) for $n=1.08$. $U=6t$ in all cases. 
HC and SQ stand for the honeycomb (with $t'=0$) 
and the square lattices, respectively. 
The temperature at which $\lambda=1$ is the $T_c$.
(b) $T_c$ as functions of the level offset $\Delta$ 
obtained by FLEX+Eliashberg equation for 
the square lattice with $U=6t$, $n=0.7$ (dashed) or for the HC 
lattice with $U=6t$ and $n=1.08$ (solid).}
\end{center}
\end{figure}

\begin{figure}[t]
\begin{center}
\includegraphics[width=8cm]{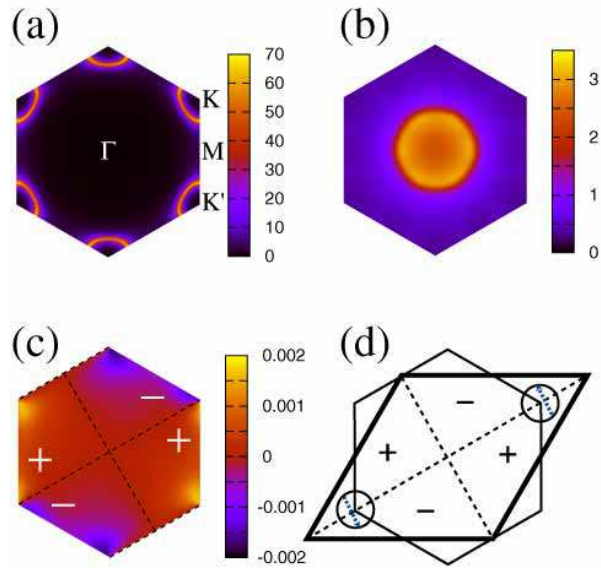}
\caption{\label{fig6} (color online) The contour plots of the FLEX result 
at the lowest Matsubara frequency for 
the honeycomb lattice ($t'=\Delta=0$) in the hexagonal Brillouin zone
with $n=1.08$ and $T=0.01t$ (a) The 
Green's function of the upper band squared,  
(b) the spin susceptibility, (c) the superconducting gap of the 
upper band,  (d) the Fermi surface (the two circles) 
and the sign of the gap function are 
schematically shown in the extended zone scheme. The dashed arrows 
represent the wave vectors at which the spin fluctuations develop.}
\end{center}
\end{figure}

The high $T_c$ obtained in this model can be understood as a 
combination of several features of this model.
Let us start with a canonical HC lattice, 
where $t'=0$ and $\Delta=0$.
We show in Fig.\ref{fig5}(a) the FLEX result of 
the maximum value of the spin 
susceptibility as a function of temperature 
for the band filling of $n=1.08$. 
The spin susceptibility 
is nearly independent of $T$,\cite{STAM,Kasahara} which is in sharp contrast 
with the Hubbard model on the square lattice.
For example, for the 
square lattice with $n=0.7$,
we have a strong enhancement of the spin susceptibility upon 
lowering the temperature. 
With further hole doping to $n=0.65$, the spin susceptibility 
is suppressed, but even there, 
the spin susceptibility moderately increases upon lowering the temperature.
Also in Fig.\ref{fig5}(a), 
we show the eigenvalue $\lambda$ 
of the linearized Eliashberg equation as functions of 
temperature for the HC and square lattices. 
$T_c$ is the temperature where $\lambda$ reaches unity. 
The DOS at 
$E_F$ is nearly the same for the square lattice with $n=0.65$ and the 
HC lattice with $n=1.08$, and 
also the spin susceptibility has similar values at low temperature, 
but still, the HC lattice has higher $T_c$.\cite{Baskaran2} 

So what is the origin of this high $T_c$ ?
Fig.\ref{fig6}(a) shows the contour plot of the Green's function squared, 
whose ridge corresponds to the Fermi surface, which 
consists of disconnected 2D pockets centered around K and K' points. 
The spin susceptibility in Fig.\ref{fig6}(b) 
is maximized at wave vectors that bridge the 
opposite sides of each pieces of the Fermi surface. 
As can be seen in Fig.\ref{fig6}(c) (and more clearly in (d))
the gap has a $d$-wave form, where the gap changes sign across the 
wave vector at which the spin susceptibility is maximized.
In contrast to the case of the square lattice, one of the 
nodes of the $d$-wave gap does not intersect the Fermi surface 
because of its disconnectivity, i.e., the gap is like ``$p$-wave'' 
if we focus only on one of the Fermi surfaces.
Since smaller number of gap nodes on the Fermi surface is 
favorable for superconductivity,\cite{KA} 
this can be a reason for high $T_c$ 
despite the low DOS and the weak spin fluctuations.

We now introduce the M-N level offset $\Delta$ 
as in the model for $\beta$-MNCl.  
As shown in Fig.\ref{fig5}(b), 
we find that superconductivity is relatively robust against the introduction 
of $\Delta$. This again is in sharp contrast with the square 
lattice, where finite $\Delta$ strongly suppresses 
superconductivity. The weak effect of $\Delta$ 
may be because the DOS of the HC lattice is 
small near the band center, so that the introduction of $\Delta$, 
which opens up a gap at the center of the band, is less effective than 
in the case of the square lattice, where the DOS 
diverges at the band center for $\Delta=0$. 

Let us now comment on the origin of the difference between the 
single and two band models, namely, low $T_c$ triplet pairing in the 
former model and the high $T_c$ singlet pairing in the latter.
The difference mainly comes from the difference in the 
band filling, i.e., for the single band model, the band filling is 
dilute, favoring ferromagnetic spin fluctuations, 
while in the two band model, it is near half filling in favor of 
antiferromagnetic spin fluctuations. 
Of course, if the level offset $\Delta$ is large enough, the 
two band case should tend to the single band case, as have been 
shown in ref.\cite{KKA}, but in the present case, 
the magnitude of $\Delta$ is small enough to retain the 
nature of the $\Delta=0$ situation.

\subsection{Gap function}
\begin{figure}[t]
\begin{center}
\includegraphics[width=6.5cm]{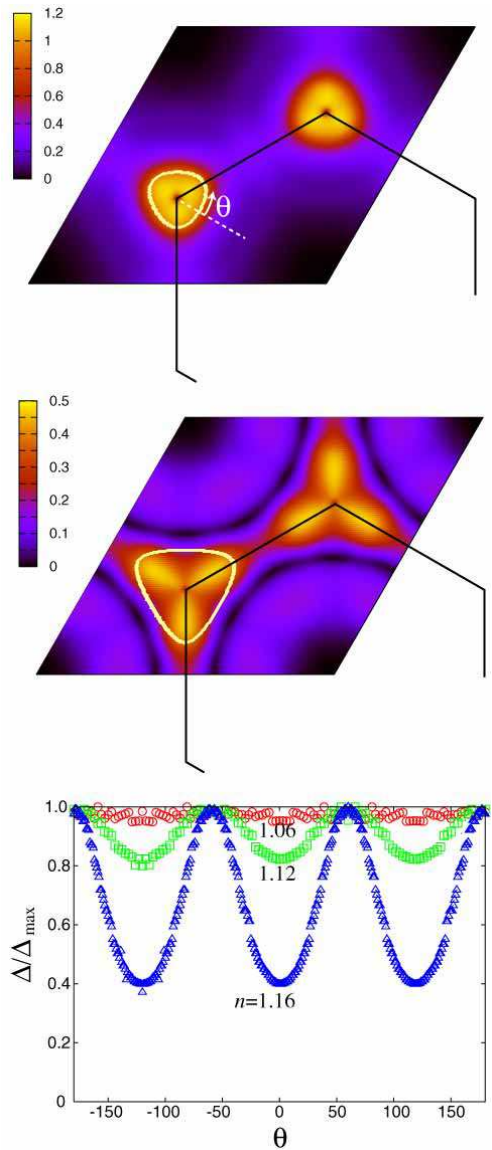}
\caption{\label{fig7} (color online) Upper two panels : 
the contour plots of the $d+id'$ gap for $n=1.06$ (top) and 
$n=1.16$ (middle). The model for $\beta$-MNCl is used.
The yellow (light) solid line (closed around the 
Brillouin zone edge) represents the Fermi surface. 
Lower panel : the normalized 
$d+id'$ gap along the Fermi surface. $\theta$ is defined in the top panel.
}
\end{center}
\end{figure}

Apart from the $T_c$, another important issue is the form of the 
superconducting gap.
By symmetry, there are two degenerate $d$-wave gaps,  
and the most probable form of the gap 
below $T_c$ is the form $d+id'$, where the two $d$-wave gaps mix with a 
phase shift of $\pi/2$.\cite{Baskaran,Kumar,Ogata} 
Since the two $d$-wave gaps have nodal lines at 
different positions, this kind of mixture leads to a gap that has a 
finite absolute value on the entire Fermi surface. 
The $d+id'$ form of the gap can be constructed from the $d$-wave gap obtained 
by solving the linearized Eliashberg equation.
In the upper panels of Fig.\ref{fig7}, we show the contour plot of the 
$d+id'$ gap for the case of $n=1.06$ and $n=1.16$ for the 
model for $\beta$-MNCl.
To clearly see the anisotropy on the Fermi surface, we plot 
the magnitude of the gap (normalized by the maximum value) in 
the lower panel of Fig.\ref{fig7}.
It can be seen that the $d+id'$ gap is 
nearly isotropic  for low doping concentration, while it becomes 
more and more anisotropic for higher doping.
The $d+id'$ pairing 
scenario is consistent with the experimental findings that the 
superconductivity occurs in the 
spin singlet\cite{Tou5}, $s$-wave like channel,\cite{Ekino,Taguchi,Kasahara} 
while the coherence peak in the spin lattice relaxation rate is 
absent.\cite{Tou4} It is also consistent with the experimental 
finding that the gap anisotropy increases upon doping.\cite{Kasahara,Hiraishi}
Since the $d+id'$ state breaks time reversal symmetry, it is 
interesting to experimentally investigate such a 
possibility.

\subsection{Effect of dimensionality}

\begin{figure}
\begin{center}
\includegraphics[width=8cm]{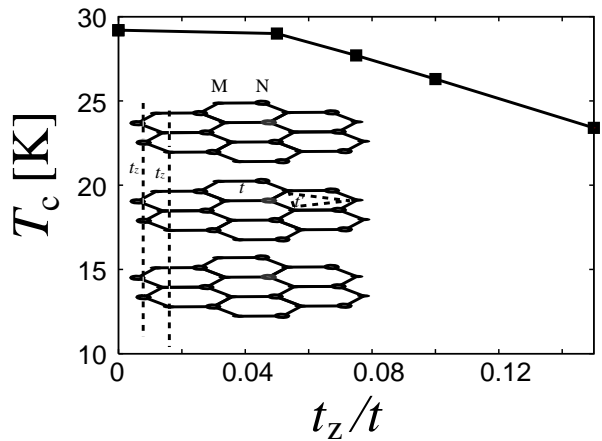}
\caption{\label{fig8} $T_c$ plotted as a function of $t_z$ for the 
model shown in the inset. $t=1.2$eV, $t'/t=0.35$, $\Delta/t=2.7$, $U/t=6$,  
$n=1.08$.}
\end{center}
\end{figure}

Finally, we consider the effect of dimensionality by adopting 
a three dimensional model where the two band model on the 
HC lattice is connected by vertical (M-M and N-N) 
hopping integrals $t_z$ (inset of Fig.\ref{fig8}). Here, 
we take up to $32\times 32\times 32$ $k$-point 
meshes and up to 8192 Matsubara frequencies. 
In Fig.\ref{fig8}, we plot $T_c$ as a function of $t_z$ for the band filling 
of $n=1.08$. 
As can be seen, the three dimensionality suppresses $T_c$, 
which is in agreement at least qualitatively with 
the experimental finding that the intercalation of THF molecules 
between the layers enhances $T_c$.\cite{Takano}

In fact, the suppression of $T_c$ with increasing three dimensionality 
is a general trend for spin fluctuation mediated pairing as has 
been studied in ref.\cite{Arita,Monthoux}. 
Namely, since the pairing interaction is large 
around a certain wave vector in spin fluctuation mediated pairing, 
the fraction of the volume where the 
pairing interaction is large in the Brillouin zone becomes smaller 
as the three dimensionality increases. 

\section{Conclusion and future problems}

In the present study, we have studied two kinds of model 
for the the doped $\beta$-MNCl. We have found that the single band model 
does not suffice as a minimal effective model of the material.
On the other hand, from the study on the two band model, 
we have proposed a possibility of spin fluctuation 
mediated $d+id'$-wave superconductivity in the doped $\beta$-MNCl.
We have shown that the relatively high $T_c$ is obtained 
as a characteristic feature of the spin fluctuation 
mediated superconductivity in a doped band insulator on the 
HC lattice.
We have also found that the gap anisotropy on the Fermi surface 
strongly increases upon increasing the doping concentration, and the 
inter layer hopping suppresses superconductivity. These results are 
in qualitative agreement with the experimental findings.

One remaining issue is the curious doping dependence of 
the superconductivity. For Li$_x$ZrNCl, 
$T_c$ shows an increase upon lowering the carrier concentration 
until a sudden disappearance of the $T_c$ and a superconductor-insulator
transition is observed.\cite{Taguchi3} Recently, this  
increase of $T_c$ has been shown to be correlated with the increase of the 
uniform spin susceptibility, which can be 
considered as a support for the present 
spin fluctuation scenario.\cite{Kasahara} In fact, although the 
spin fluctuation is nearly {\it temperature} independent in our model
(so that the development of the spin fluctuation upon lowering the 
temperature need not be observed experimentally), 
it is strongly {\it doping} dependent and is enhanced upon lowering the doping 
rate.\cite{Kasahara}
Nonetheless,
the origin of the {\it sudden} drop of $T_c$ at a certain doping rate 
can not be explained within the present two band 
model (see Fig.\ref{fig4}) and 
remains as an open problem. 
Also, the almost constant behavior of $T_c$ in the heavily 
doped regime\cite{Takano} is puzzling since for such a large doping, 
the Fermi level hits the second band 
and a large difference in the DOS should be detected 
(which is not the case experimentally). A rigid band 
picture may not be valid in the heavily doped regime, 
as suggested in some experiments\cite{Yokoya}
and band calculations.\cite{Oguchi2}

The author acknowledges Y. Taguchi, Y. Iwasa, Y. Kasahara, H. Tou, 
R. Kadono, M. Hiraishi, S. Yamanaka, T. Ekino, A. Sugimoto, H. Fukuyama, 
T. Oguchi, M. Onoue, and H. Aoki for fruitful discussions.
Numerical calculation has been done at the computer center, ISSP, University 
of Tokyo. This study has been supported by Grants-in Aid from MEXT of Japan 
and JSPS.


\begin{thebibliography}{9}
\bibitem{comment} Generally, the on-site repulsion on the 
M site and N site can be different, but here we assume them to be 
equal for simplicity.
\end{thebibliography}
\end{document}